\documentclass[reprint,groupedaddress]{revtex4-1}

\usepackage[T1]{fontenc}       % Use modern font encodings
\usepackage{amsmath}
\usepackage{color}% only for revision, suppress at end
\usepackage{graphicx}

\newcommand{\degrees}{^{\circ}}

\begin{document}
%   TITLE

\author{Christophe E. Wilhelm}
\affiliation{School of Electrical and Electronic Engineering, Nanyang Technological University, Singapore 639798 Singapore}
\affiliation{CINTRA (CNRS International  NTU THALES Research Alliance), Singapore 639798 Singapore}
\affiliation{Thales Research and Technology, 91767 Palaiseau France}
\author{M. Iqbal Bakti Utama}
\affiliation{Division of Physics and Applied Physics, School of Physical and Mathematical Sciences, Nanyang Technological University,
Singapore 637371 Singapore}
\author{Qihua Xiong}
\affiliation{Division of Physics and Applied Physics, School of Physical and Mathematical Sciences, Nanyang Technological University,
Singapore 637371 Singapore}
\author{Cesare Soci}
\affiliation{Division of Physics and Applied Physics, School of Physical and Mathematical Sciences, Nanyang Technological University,
Singapore 637371 Singapore}
\affiliation{CINTRA (CNRS International  NTU THALES Research Alliance), Singapore 639798 Singapore}
\author{Ga\"{e}lle Lehoucq}
\affiliation{Thales Research and Technology, 91767 Palaiseau France}
\author{Daniel Dolfi}
\author{Alfredo De Rossi}
\email{alfredo.derossi@thalesgroup.com}
\author{Sylvain Combri\'{e}}
\affiliation{Thales Research and Technology, 91767 Palaiseau France}

\title[Nanowire on $SiN_x$ PhC]  {Broadband tunable hybrid photonic crystal-nanowire light emitter}

\begin{abstract}
We integrate about 100 single Cadmium Selenide semiconductor nanowires in self-standing Silicon Nitride photonic crystal cavities in a single processing run. Room temperature measurements reveal a single narrow emission linewidth, corresponding to a Q-factor as large as 5000. By varying the structural parameters of the  photonic crystal, the peak wavelength is tuned, thereby covering the entire emission spectral range of the active material. A very large spectral range could be covered by heterogeneous integration of different active materials.
\end{abstract}

\maketitle
%\makeatother

% \includegraphics[width=0.6\textwidth]{fig_abstract_test}
 %
The miniaturization of laser devices\cite{hill_advances_2014} is crucial in the context of on-chip optical communications, but also for medical imaging or sensing \cite{he_detecting_2011}. Furthermore, scaling down the size of the laser helps in reducing the power consumption through the reduction of the active volume, a critical issue where a massive number of emitters is meant to coexist in a tiny space. A third implication is an intrinsically faster response\cite{hill_advances_2014}.  Since the available gain is reduced when down-sizing the active volume, it is crucial to associate the emitter to a small but high quality optical cavity\cite{couteau_nanowire_2015}.\\
In surface-plasmon based cavities the optical confinement breaks the diffraction limit, hence allowing an  extremely high enhancement of the light-matter interaction. However, in this regime the optical field oscillates by periodically exchanging its energy with electrons in the metal, where Joule effect induces a very fast decay\cite{khurgin_how_2015}. This is not an issue in the context of fast extraction through enhanced stimulated emission\cite{ji_non-blinking_2015}, but it compromises the goal of energy-efficient light emission, particularly in the visible and near infrared spectral domain.\\ 
Optical confinement, approaching the diffraction limit is possible in dielectric cavities based on the Bragg scattering, e.g. Photonic Crystals (PhC), still keeping optical losses to a very low level\cite{zhang_small-volume_2004}. Thus, continuous wave laser operating at room temperature with a bias current of tens of $\mu A$s has been demonstrated by combining a very small active area with a high-quality PhC cavity\cite{takeda_heterogeneously_2015}.  These lasers have been hetereogenously integrated on a silicon chip and the technology is, in principle, able to address the challenge of massive integration.\\
The concept of hetereogeneous integration can be pushed further so that the active material, e.g. a nanowire (NW) or a colloidal quantum dot, which are synthesized in large number, is integrated into a passive dielectric structure and, ultimately, in a complex photonic circuit. A single III-V nanowire has been inserted into a Silicon PhC waveguide, which creates an optical cavity self-aligned with the emitter\cite{birowosuto_movable_2014}. The heterogeneous integration on silicon chip is an important result. Also, the complexity and the criticality of the manufacturing process are reduced, as the fabrication of the active material and of the photonic chip are separate.\\
In this paper, we create a self-aligned cavity by inserting a single CdSe nanowire into a Silicon Nitride Photonic Crystal.  By mapping the NW positions by topographic Atomic Force Microscopy measurements, and accurately constructing PhCs around the NWs rather than manually aligning them individually, we demonstrate fabrication of about one hundread hetereogeneous devices at the same time, with substantially equivalent optical functionality.\\

\begin{figure*}
\includegraphics[width=0.7\textwidth]{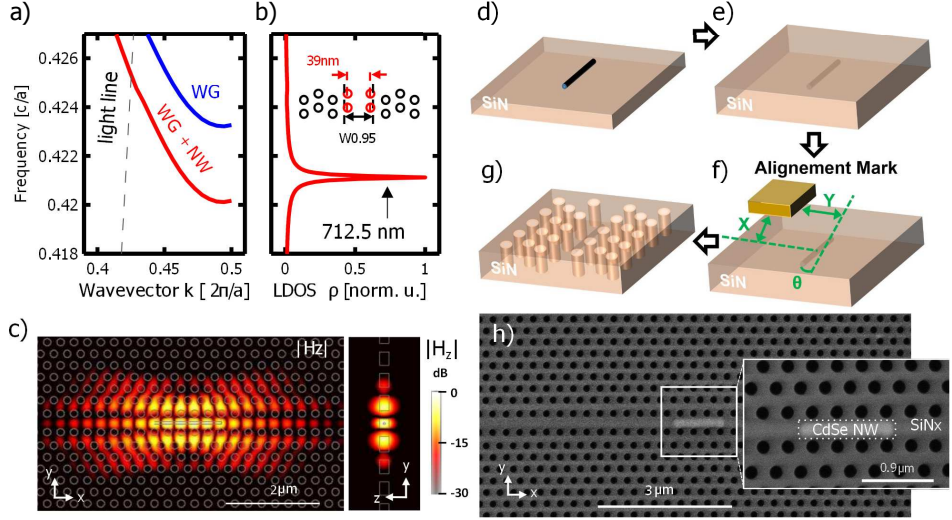}
\caption[fig1]{\label{fig1}  Concept of the PhC-NW cavity: calculated dispersion of the PhC waveguide with (red) and without NW (blue) a) and, b), Local Density of the States (normalized) peak (Q=5300); the inset represents the design of the PhC WG; distribution of the magnetic field $|H_{z}|$ at the resonance, represented as a false-color map in logarithmic scale c). Fabrication process flow of the PhC-NW source d-g). Deposition of $SiN_{x}$ on the substrate d); NW dispersion and deposition of the $SiN_{x}$ cladding layer e); Localization of the NW by AFM f); patterning of the PhC aligned to the NW g) prior to the local removal of the substrate; Scanning Electron Micrograph from the top of the PhC suspended membrane with a magnified view in the inset h).}
\end{figure*}

The device concept is shown in Fig.\ref{fig1}. The starting point is a PhC waveguide made of Silicon Nitride ($SiN_x$), which is characterized by a fundamental mode with the dispersion shown in Fig.\ref{fig1}a. The addition of a NW along the axis of the waveguide increases the effective refractive index hence it red shifts the dispersion. This creates an effective potential well for the optical field which allows a resonance (Fig \ref{fig1}b), spectrally centered just between the edges of the two bands in Fig \ref{fig1}a. The corresponding field distribution is well localized around the nanowire (Fig.\ref{fig1}c), in spite of the weaker index contrast available.\\
The structure is a self-standing 270 nm thick membrane of $SiN_x$, patterned with a hexagonal lattice with constant $a=300$ nm of holes  with radius $r=83$ nm.  A \textit{line defect} of missing holes induces a waveguide. Further modifications of the lattice are made to tailor the field confinement and to ensure that the propagating mode is not leaky (since their dispersion is below the light line). These changes are: a dislocation of the two half-lattices such that their relative distance is $W=0.95\sqrt{3}a$ and inward shift of first line of holes by 39 nm (Fig.\ref{fig1}b, inset). This fine tuning is crucial in order to cope with the substantially smaller refractive index of $SiN_{x}$  ($n=1.9$) relative to semiconductors (e.g. Silicon $n=3.4$), which considerably restricts the parameter space in the design.  This is a well know issue when creating PhC optical nanostructures using low-index materials\cite{barth_modification_2007,bayn_ultra_2008}.\\
The choice of this material is however motivated by its broad transparency range, extending to the visible and near UV spectral range owing to its large band gap ($>4$eV)\cite{milek1971handbook,Zanatta2013} and by the fact that it can be deposited as a thin film, implying flexibility and easy integration of heterogeneous materials. It is also important to note that $SiN_{x}$ is compatible with biological tissues\cite{gao_bioengineering_1997}.\\
The calculations in Fig.\ref{fig1}a-c consider a CdSe NW (refractive index 2.85) with diameter 60 nm and length 1.5 $\mu m$. These values result from an optimization. Indeed, thicker NWs would induce a deeper potential well, resulting into a more abrupt confinement and, hence, a much stronger optical leakage\cite{akahane_high-q_2003}, substantially decreasing the Q-factor.  On the other hand, the yield of thinner NWs would decrease very fast, also because of the increased role of non-radiative surface recombination. Furthermore, longer NWs tend to bend during deposition and, therefore, are more difficult to process. Similar considerations are valid for NWs made of other materials, e.g. GaAs.\\
The calculations are made by our in-house developed 3D Finite Differences in Time Domain (FDTD) code ensuring relative accuracy in the calculated frequency within $\frac{\Delta f}{f}=10^{-3}$. The resonance corresponding to a localized mode appears in the plot of the (normalized) Local Density of Optical States (LDOS), which is proportional to the spontaneous emission, calculated following Ref.\cite{Xu1999}. We approximated the source as a dipole located at the center of the NW and oriented along $y$. 
The Q-factor of the resonance is 5300 (i.e. linewidth = 0.13 nm), while the corresponding wavelength is 712 nm, well within Photo Luminescence (PL) spectrum of the CdSe NWs \cite{chen_excitonic_2011}. 
The  mode volume $V=8.5\times10^{-20}m^{3}=2(\frac{\lambda}{n})^{3}$, i.e. reasonably close to the diffraction limit $\left(\frac{\lambda}{2\,n}\right)^3$. Hence, the overlap of the field with the NW is $\Gamma=V_{NM}V^{-1}=0.05$, which is good, considering the small size of the emitter. \\
%
%
%              FIGURE 1
The fabrication of the structure is shown in  Fig.\ref{fig1}d-f. First, about half $SiN_x$ layer (120 nm) is deposited on a <001> Gallium Phosphide (GaP) substrate\cite{note_GaP}, by high temperature (360$\degrees\mathrm{C}$) and low mechanical stress Plasma Enhanced Chemical Vapor Deposition (PECVD). High quality single phase Wurtzite CdSe nanowire are grown using a CVD process described in Ref.\cite{utama_vertically_2011} with van der Waals epitaxy on muscovite mica substrate.\cite{utama2013}.  They are then removed from their growth substrate by placing them in an ethanol solution and then in an ultrasonic bath for a few seconds \cite{patolsky_fabrication_2006}. The solution is then dropped on the $SiN_{x}$ layer and the solvent is removed by heating at $110\degrees\mathrm{C}$.\\
The NWs are further cladded by a 150 nm thick layer of $SiN_{x}$ (Fig.\ref{fig1}e), which allows to protect the NW and to symmetrize the structure. At this stage, we haven't taken any specific measure to control the position of the NWs, although we point out that several techniques have been proposed for the transfer and the deterministic placement of NWs on a planar substrate, for instance using the Langmuir-Blodgett technique\cite{whang_large-scale_2004}. Thus,  the buried NWs are spatially mapped by detecting the induced relief at the surface using AFM (scanning rate 0.1 Hz, in-plane resolution 22 nm/pixel) on an area encompassing alignment marks and several NWs. \\
The position and the angle of each NW are retrieved relative to the alignment marks (Fig.\ref{fig1}f). Only a subset of the detected NW is retained for the fabrication of the device, according to their diameter,  $50nm\pm15nm$ (out of a broader distribution: 35 to 130 nm) and their length,  chosen to be close to 1.5 $\mu m$, since longer wires tend to bend during evaporation.\\
Then, PhC are fabricated following a fairly standard sequence of positive resist (PMMA-A4, thickness 200 nm) exposure by electron beam lithography (NANOBEAM NB4), and reactive ion etching of the $SiN_{x}$ layer (Fig.\ref{fig1}g) using a $CHF_{3}/O_{2}$ mixture \cite{gatzert_investigation_2006}.
The last step is the wet chemical etching of the GaP substrate\cite{plauger_controlled_1974} under the $SiN_{x}$ Photonic Crystal in order to obtain a self-standing membrane, providing the required refractive index contrast  below the structure.\\
The finalized device is shown in Fig.\ref{fig1}h. As apparent in the inset, the alignment of the NW with the PhC is very good. About one hundred of these have been fabricated in a single run. On average over the investigated devices, the positioning and angular errors are 53 nm and $0.1\degrees$ respectively.\\
%
%
%              FIGURE 2
\begin{figure}
\includegraphics[width=1\columnwidth]{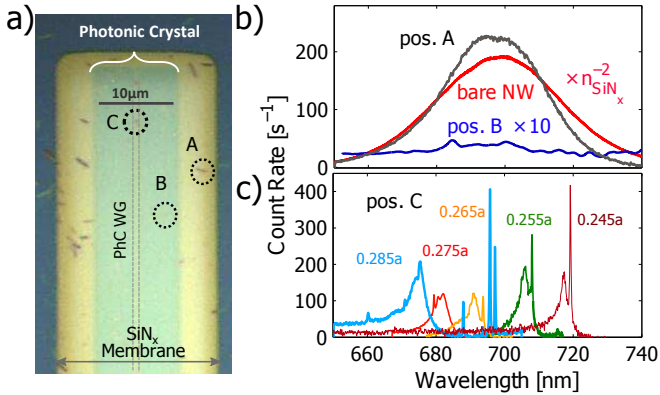}
\caption[fig2]{\label{fig2} Photoluminescence of randomly distributed NWs. a)  optical image of the sample, circles point to the NWs whose PL is shown in panels b and c; b) NW in homogeneous $SiN_{x}$ (gray, position \textbf{A}) and patterned $SiN_{x}$ (blue, position \textbf{B}, multiplied by 10) and reference
NW on a Quartz substrate (red); c) PL when NW crosses the PhC waveguide (position \textbf{C}), color codes relate to the reduced PhC hole radius, the width of the twin narrow peaks (cyan line) is about 300 pm. The vertical axis in panels b and c can be compared quantitatively.}
\end{figure}
%              FIGURE 2
%
Before considering the emission properties of these structures, and in order to appreciate the benefit of spatial alignement, it worth investigating devices fabricated following the same procedure, except that NWs are not aligned into the PhC cavity (e.g. the NW labelled as A in  Fig.\ref{fig2}a). The micro-Photo Luminescence (PL) is performed using a commercial equipment  (Renishaw), under continuous wave pumping using the 514 nm line of the Argon laser. A nearly diffraction limited spot size is obtained using a $63\times$ and $N.A.=0.95$ microscope objective.  The pump is circularly polarized, in order to remove the sensitivity of the measurement to the absorption anisotropy of the NWs.\cite{Wilhelm_Anisotropy_2012,Ruda2005}\\
First, we point out that embedding the NWs in $SiN_{x}$  does not alter their emission properties substantially (beside the trivial effect on the collection efficiency). This is apparent in Fig.\ref{fig2}b, where the PL spectra of the NW located outside the PhC area (labeled as \textbf{A} in Fig.\ref{fig2}a) and of the reference NW lying on a Quartz substrate are very similar, both centered at about 700 nm and about 30 nm broad. 
Although an accurate quantitative comparison is not easy, because the PL yield varies from NW to NW, it is apparent that the PL is strongly reduced by about 2 orders of magnitude, when the NW is located inside the PhC (position labeled as \textbf{B}). This is way larger than any possible fluctuation in the PL of different NWs. This result is reproduced systematically and is consistent with established literature on the inhibition of the spontaneous emission in PhC\cite{fujita_simultaneous_2005}.\\
The PL spectrum is drastically different when the NW crosses the PhC waveguide (Fig.\ref{fig2}c), revealing peaks which are correlated with the radius of the PhC holes as expected (the larger the radius, the shorter the wavelength). Importantly, the PL is comparable in intensity as in the case of NWs embedded into a homogeneous $SiN_x$ layer.\\
The strong modification of the PL spectra of emitters embedded in PhC waveguides is due to the strongly dispersive and large LDOS \cite{Lecamp_Spontaneous}. The narrow peaks ($\Delta\lambda\approx$ 300 pm) correspond to the defect modes (lower in energy) induced by the NW in the PhC waveguide. Thus, the perturbation is enough to induce a resonant mode with spectral position mainly dictated by the PhC geometry, even if  the position of the NW relative to the waveguide is not controlled.

%              FIGURE 3
\begin{figure}
\includegraphics[width=1.0\columnwidth]{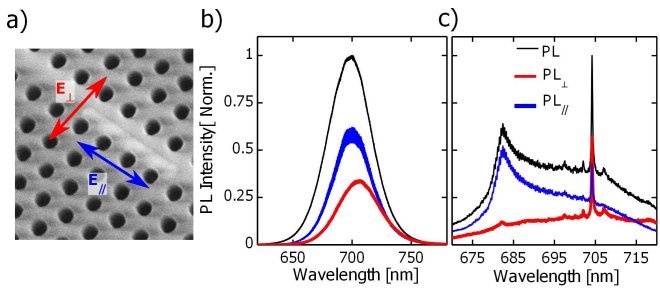}
\caption[fig3]{\label{fig3} Identification of the cavity mode. a) SEM image of the PhC cavity (radius $0.285a$) aligned to a NW (sample 5). The parallel ($//$) or perpendicular ($\perp$) alignment of the analyzer is referred to the waveguide axis; b) PL of reference NW on the Quartz substrate; c) PL of the NW 5, color code is the same as in b) and refers to the polarization.}
\end{figure}
%              FIGURE 3
%
%
The PL measurements on aligned NWs, obtained from the full fabrication process as described above, is shown in Fig.\ref{fig3}. In order to formally associate the expected resonant peak to the designed cavity mode, we also consider the polarization of the PL. While the emission of bare NWs on a Quartz substrate is moderately anisotropic (Fig.\ref{fig3}b), with a larger parallel term\cite{Wilhelm_Anisotropy_2012,Ruda2005,Mishra2007}, the structure of the PL emitted by our device reveals a broad and a narrow peaks.  The broad peak, located at higher energies, inherits the polarization properties of the emitter, and might be related to the higher order mode of the PhC waveguide. The narrow peak is mainly polarized perpendicular, as expected from the calculation of the radiated Far Field of the cavity mode represented in Fig.\ref{fig1}c.
Moreover, the measured and calculated Q factor and peak wavelength are very close.
We point out that the relative strength of the two peaks is not entirely representative of the power radiated. In fact, PhC cavity mode radiates mainly at a grazing angle with the surface of the sample, and, therefore, only a small part of the power emitted is collected by the microscope objective. In contrast, the background PL is radiated much more isotropically, and, therefore, collected more efficiently. Note that it is possible to improve substantially the directivity of the radiation from PhC cavities (and thereby the collection nearly to 50\%) by a controlled modification of the design \cite{Tran2009,Tran2010}.\\
%
 %              FIGURE 4 (ex figure 5)
\begin{figure*}
\includegraphics[width=0.55\textwidth]{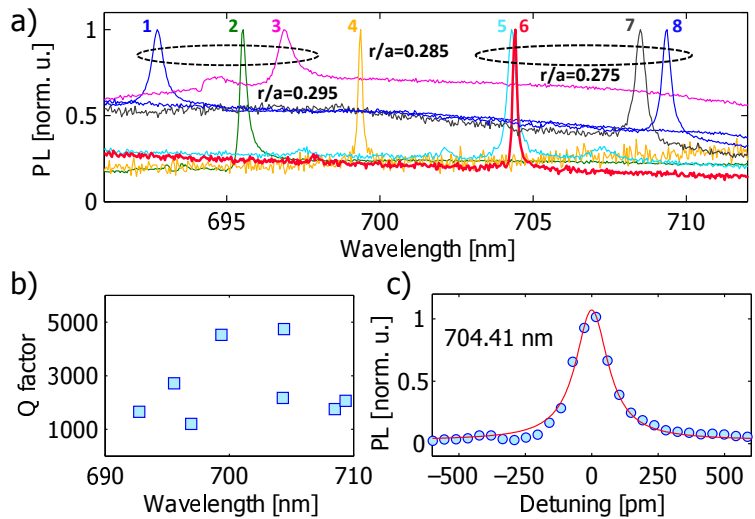}
\caption[fig4]{\label{fig4} Comparison of the emission properties of sample 1 to sample 8. a) ensemble view, dashed circles denotes a common radius for the PhC holes; b) spectral width of the peak (Q factor); c) detailed lineshape and Lorentzian fit (sample 6). The PL is collected perpendicular to the NW axis.}
\end{figure*}
%
%              FIGURE 4
%
Here, is is shown one out of the eight devices which have been selected for systematic measurements, numbered from 1 to 8 according to the increasing peak wavelength. The same polarization properties have been observed for  the others, as expected. We focus now on the properties of the cavity mode and therefore, we plot the perpendicular component of the spectra in Fig.\ref{fig4}a.  These 8 peaks relate to three different design (hole sizes) of the PhC and, indeed, they are clustered accordingly. There is an apparent dependence on the NW, depending primarily on its size (diameter between 30 and 90 nm) and secondly on the accuracy of its positioning (within 60 nm, except for sample 7). Interestingly, in sample 7 the NW is misplaced by exactly one lattice period, and still the PL yield is strong. This is not surprising given the spatial structure of the cavity mode (Fig.\ref{fig1}c). The measured Q-factors range from 1000 to 5000 (Fig.\ref{fig4}b,c) and match well to a Lorentzian lineshape (Fig.\ref{fig4}c). The highest Q factors are obtained for the NWs 4 and 6 (the thinnest NWs). It is also interesting to note that best placement (error is 25 nm) is achieved for sample 6. These results are consistent with FDTD calculations, predicting a monotonous decrease of the Q factor and a red shift as the diameter of the NW increases from 30 nm to 90 nm. Also a positioning error larger than about 30 nm induces a decrease of the Q factor and a red shift.
We do not believe that the absorption properties of the NW itself plays a role on the spectral width of the peaks, although we cannot rule out it.\\
Figure \ref{fig4} substantiates the main point of this letter, namely, that is is possible to generate a variety of  wavelengths on the same chip by design. This is a fundamental property either for communications (wavelength domain multiplexing) or for labeling in biologic detection.
%
 %              FIGURE 5
\begin{figure*}
\includegraphics[width=0.6\textwidth]{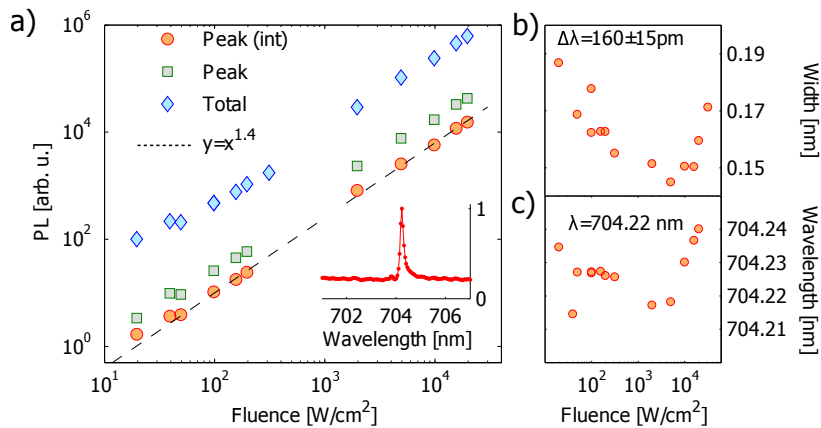}
\caption[fig5]{\label{fig5} PL dependence on the pump fluence, NW 6. a) total integrated PL spectra (diamonds), peak (squares) and integrated peak (circles), dotted line is the $y=x^{1.4}$ law as a guide for the eyes, spectra is in inset; b) spectral width of the peak (FWHM) and its average; c) peak spectral position and its average. PL is collected perpendicular to the NW axis.}
\end{figure*}
%              FIGURE 5
The PL spectra and specifically the width of the peak do not reveal any substantial change with the intensity of the excitation\cite{note_spot_area}. Specifically, the ratio between the total PL and the peak intensity is constant over more than three decades (Fig.\ref{fig5}a), indicating that the PL signal is entirely due to spontaneous emission. 
Interestingly, the dependence on the PL on the pump is superlinear, following a law $PL\propto I^{1.4}$. In another sample we have observed a similar behavior $PL\propto I^{1.65}$. We believe this is entirely related to the NWs, as a very similar behavior has been  been observed in CdSe NW under the same conditions (RT and CW pumping)\cite{Vietmeyer2011} and explained as follows: at intermediate pumping levels (which covers our parameter space entirely), it is assumed that the hole population depends linearly on the pumping rate, because of fast hole trapping, while electrons have the usual dependence on the square root of it. This results into a theoretical $PL\propto I^{1.5}$ dependence of the PL, which is consistent with our findings.\\
We have pumped the NW at the maximum fluence level of 20 $kW cm^{-2}$, which is likely to be still below the transparency threshold of the NWs. We estimate this threshold based on recent experiments on one ensemble of similar NWs\cite{chen_excitonic_2011}. There, random lasing has been observed with a minumum threshold of about $200 kW cm^{-2}$. Consistently, we performed additional measurements on a single bare NW (on Quartz)  and observed the onset of stimulated emission at a fluence of $100 kW cm^{-2}$.\\
Lasing of on fairly larger  (200 to 500 nm) single CdSe NWs\cite{xiao_single-nanowire_2011} has been observed at lower pump levels. This corroborates the hypothesis that surface recombination degrades the PL emission at Room Temperature substantially in our thin NWs (which is also apparent in the power dependence of the PL signal). The use of core-shell structures, such as AlGaAs/GaAs\cite{mayer_lasing_2013,saxena_optically_2013}, or GaAs/GaP\cite{priante_abrupt_2015} could greatly help in achieving efficient lasing. Also, the telecom spectral domain could be addressed by suitable choice of the III-V alloy. Interestingly, the beta factor of our device is estimated based on the mode volume to be about 0.1, using formula in Ref. \cite{larrue_monolithic_2012}. This is a typical number for nanolasers. The required material gain at threshold is estimated to about 300 $cm^{-1}$, which is well within reach.\\
%
%CONCLUSIONS
%
In conclusion, we have demonstrated a novel approach to the heterogeneous integration of semiconductor nanowires in a small volume optical cavity. The resonant optical field is automatically localized around the nanowire, resulting into an efficient coupling. The key aspect of this approach is that the cavity is fabricated after the transfer of the nanowires on the substrate and, consequently, can be accurately aligned on it. This allows for the fabrication of hundreds of samples in a single processing run. We show that the photoluminescence is controlled by the cavity mode, which allows for the generation of multiple wavelengths with fairly narrow peaks (160 pm) on the same optical chip. Lasing has not be achieved at the moderate pumping level used for characterization ($20 kW/cm^2$), but it should be within reach. Furthermore, the use of core-shell nanowire structures should allow efficient light emission, while the collection efficiency can be improved by design. For some applications, such light sources could be easily separated from their substrate and attached as wavelength-labeled markers to suitable targets.
\section{acknowledgement}
Authors are thankful to Dr. C. Couteau for valuable discussions, D. Thenot for help on the SiN depositions, S. Xavier for help on the e-beam lithography and B. Servet for his initial help on the PL set-up. Q.X. gratefully acknowledges the financial support from Singapore National Research Foundation vis NRF Investigatorship Award (NRF-NRFI2015-03), and Ministry of Education via Three Tier2 grants (MOE2011-T2-2-051,  MOE2012-T2-2-086 and MOE2013-T2-1-049). C.S. acknowledges financial support of the Singapore Ministry of Education (MOE2013-T2-1-044).
\bibliography{PhC_nanowire_CW}

\begin{thebibliography}{37}
\bibitem [{not({\natexlab{b}})}]{note_spot_area}%
  \BibitemOpen
  \href@noop {} {\enquote {\bibinfo {title} {the estimated area for the
  focusing spot is $1.5\mu m^{2}$},}\ } ({\natexlab{b}})\BibitemShut {NoStop}%
\end{thebibliography}
 
\end{document}